\documentclass[aps,prl,twocolumn,showpacs]{revtex4}
\usepackage{eurosym}
\usepackage{amsfonts}
\usepackage{amsmath}
\usepackage{amsopn}
\usepackage{amssymb}
\usepackage{graphicx}
\usepackage{graphics}
\usepackage{color}
\usepackage{verbatim}

\newcommand{\tmu}{\tilde{\mu}}

\newcommand{\tu}{\tilde{u}}

\newcommand{\tbPsi}{\tilde{\bf \Psi}}

\newcommand{\br}{{\bf r}}

\newcommand{\bL}{{\bf L}}

\begin{document}

\title{Non-coaxial vortices in two-component Bose-Einstein condensates:
Persistent precession and nutation}
\author{R. Driben$^{1,2}$, V. V. Konotop$^{3}$, and
	T. Meier$^{1}$}
\affiliation{$^1$Department of Physics and CeOPP, University of Paderborn, Warburger Str. 100, D-33098 Paderborn, Germany
	\\
$^{2}$ITMO University, 49 Kronverskii Ave., St. Petersburg 197101, Russian Federation
\\
$^3$Centro de F\'isica Te\'orica e Computacional and Departamento de F\'isica, Faculdade de Ci\^encias, Universidade de Lisboa, Campo Grande, Edif\'icio  C8, Lisboa 1749-016, Portugal }

\begin{abstract}
It is demonstrated that a two-component Bose-Einstein condensate (BEC) with all-repulsive inter-atomic interactions
loaded into a radially symmetric harmonic trap supports robust non-coaxial vortices
with approximately orthogonal vortex lines in each of the components.
These cross vortices are excited from the linear modes by a sudden switch-on of the nonlinearity (via Feshbach resonance) and are characterized by persistent dynamical regimes of precession with nutation, resembling the motion of a rigid body.
The obtained dynamics can be understood qualitatively on the basis of a simple mechanical  model.
\end{abstract}

\maketitle

Quantized vortices are fundamental topological objects responsible for a variety of interesting phenomena observable in such areas as superconductivity~\cite{superconduct},
superfluidity~\cite{superfluid}, 
and Bose-Einstein condensates (BECs)~\cite{Pitaevski,Fetter}.
In a three-dimensional (3D) one-component BEC, where  a vortex is characterized by its charge, as well as by the geometry and dynamics of the vortex line,
vortices received considerable attention during the last two decades. Multiple studies were devoted to vortex shapes and their stability~\cite{Sinha,DBEC,SvFetter,VPG,Barcelona,Kevrekidis}.
More recent investigations of BECs concentrated on the dynamics of vortex lines, in particular Kelvin waves~\cite{Kelvin}, vortex clusters~\cite{vortexclusters}, and dipoles~\cite{vortexdipoles}, as well as on the existence of more complex topological objects like hybrid vortex solitons~\cite{DKMMT} and twisted toroidal vortex solitons~\cite{KMS} in inhomogeneous media.

Even richer topological structures appear in multicomponent condensates~\cite{KaTsU}. These are, in particular, mixtures of hyperfine states, like the ones explored since the very first experiments reported observation of vortices in BECs~\cite{experim1,experim2,experim3, experim4}. Such BECs have a spinor nature and are characterized by macroscopic order parameters with two or more components.  Nowadays, the complex dynamics of the phase separation~\cite{Kverekidis_Hall} as well as sophisticated topological states 
like vortex solitons~\cite{Kevrekidis2}, skyrmions~\cite{RuAn}, vortex lattices~\cite{MiMoDu}, and Dirac magnetic monopoles~\cite{Dirac_monopoloe} in spinor BECs have been described theoretically and observed experimentally.

Here we present non-coaxial vortices existing in two-component BECs, which have a significantly simpler geometry than the topological defects mentioned above.
The objects we report here consist of two vortex lines, each one belonging to one of the components of a binary mixture (we term them {\em cross vortices}).
Families of cross vortices solutions bifurcating from two orthogonal vortices existing in the linear decoupled condensates and differ in many aspects form the most of topological structures studied, so far.
In particular, their generation does not require trap rotation or linear coupling between the components.
Such cross vortices are observed in miscible configurations~\cite{Mineev}, where each component has nonzero density both inside and outside the core region of the other component and is characterized by a hole where the vortex lines intersect. 
The reported objects can be excited by switching-on the nonlinearity and support a plethora of persistent dynamical regimes when vortices interact being oriented by arbitrary relative angles. Among all such regimes, we concentrate here on persistent precession and nutation which closely resemble
the dynamics of a heavy top~\cite{precessions}.

We consider a binary mixture of BECs of the atomic components having equal atomic masses $m$,  and loaded in radially symmetric parabolic traps, identical for both components and characterized by the linear oscillator frequency $\omega_{0}$. We
also assume that the intra-species scattering lengths  $a_{1,2}$ are equal, i.e., $a_1=a_2=a$, and the inter-species interactions are characterized by the s-wave scattering length $a_{12}$.
Further, we adopt  dimensionless units where time and coordinates are measured in  $2/\omega_0$ and $a_0=\sqrt{\hbar/m\omega_0}$ units, respectively.
The mixture is described by the two-component macroscopic wave function (spinor)  $\Psi=(\psi_1,\psi_2)^T$ which solves coupled dimensionless Gross-Pitaevskii (GP) equations
\begin{subequations}
	\label{GPE}
\begin{eqnarray}
i\frac{\partial \psi_1 }{\partial t}=-\nabla ^{2}\psi_1 + r^2\psi_1 +\left(|\psi_1|^{2}+\alpha |\psi_2|^2\right)\psi_1
\label{GPE_1}
\\
i\frac{\partial \psi_2 }{\partial t}=-\nabla ^{2}\psi_2 + r^2\psi_2 +\left(|\psi_2|^{2}+\alpha |\psi_1|^2\right)\psi_2
\label{GPE_2}
\end{eqnarray}
\end{subequations}
Here
  $\alpha=a_{12}/a\in[0,1]$ characterizes the relative strength of the inter- and intra-species interactions, i.e.,
we consider positive scattering lengths when the homogeneous mixture is miscible (a stratified phase is absent~\cite{Mineev}).
In the chosen normalization the total number of atoms is given by ${\cal N}=Na_0/(8\pi a)$ with $N=\int \Psi^\dag\Psi d\br$.


Equations (\ref{GPE}) conserve the number of atoms in each component, i.e. $N_{1,2}$, the total energy $E=E_1+E_2+E_{int} $ where $E_j=\int\left(|\psi_{j,x}|^2+r^2|\psi_j|^2+\frac 12|\psi_j|^4\right)dx$ is the energy of the $j$-th component and $E_{int}=\alpha\int|\psi_1|^2|\psi_2|^2dx$ is the energy of inter-species interactions, as well as the total angular momentum $\bL=\bL_1+\bL_2$ where $\bL_j=-i\int \psi_j^*(\br)(\br\times\nabla) \psi_j(\br)d\br$.
Meantime, neither the energies $E_{j}$ nor the angular momenta $\bL_j$ are conserved as function of time.
In particular one obtains
\begin{eqnarray}
\label{M12}
\frac{d\bL_1}{dt}=-\frac{d\bL_2}{dt}=\alpha\int|\psi_2|^2(\br\times\nabla) |\psi_1|^2d\br,
\end{eqnarray}
i.e., the change of the individual components of the angular momenta occur because of the inter-species interactions.

Usually, the generation of single vortices is achieved either by rotating traps~\cite{Fetter} or  by dynamical~\cite{experim1} and topological~\cite{topol,experim4} phase imprinting.
Nonlinear interactions of our model
(\ref{GPE})allow us to adopt the following methodology.
Initially two (linear) vortices are created in the absence of two-body interactions and thus independently in each of the components. Then the  nonlinearity is switched on (by means of the Feshbach resonance).
Even though this represents a rather strong initial "perturbation" of the nonlinear vortex states,
we demonstrate below that the objects created in such a way are robust cross vortices, which are accompanied by rotational dynamics.
We notice that growing a vortex state departing from the linear limit was previously discussed in the literature~\cite{DBEC}, however,
in the context of a one component condensate and for a gradual (rather than instant) increase of the inter-atomic interactions.

Let us now turn to the
stationary solutions of  (\ref{GPE}) with the components having equal chemical potentials $\mu$, for which $\psi_j=e^{-i\mu t}u_j({\br})$ with $u_{j}({\br})$ solving the stationary GP equations ($j=1,2$)
\begin{eqnarray}
\label{statGPE}
\mu u_j =-\nabla ^{2}u_j + r^2u_j +\left(|u_j|^{2}+\alpha |u_{3-j}|^2\right)u_j .
\end{eqnarray}
We start with the linear eigenvalue problem $\tmu\tbPsi=(-\nabla^2+r^2)\tbPsi$ (hereafter we use tilde to specify the eigenvalues and eigenfunctions of the linear limit). The components are now decoupled and the eigenstates are
given by the well known eigenmodes of the linear harmonic oscillator.
Without loss of generality, we fix the $z$-axis along the vortex line of the simplest linear vortex of the first component, i.e.,
$
\tilde{u}_1(\br)=N_1^{1/2}\pi^{-3/4}(x+iy)e^{-r^2/2},
$
which corresponds to the chemical potential $\tmu=5$ and has $N_1$ atoms (change of the direction of the vortex lines along axes will affect only the direction of the rotations discussed below). The vortex line in the second component can be taken to be rotated by an angle $\beta$ with respect to the second chosen axis.
We chose it to be the $y$-axis which
means that $\tilde{u}_2 (\br^\prime)=N_2^{1/2}\pi^{-3/4}(x^\prime+iy^\prime)e^{-r^{\prime 2}/2},$ where 
\begin{eqnarray*}
\label{R}
\br^\prime=R_x(\beta)\br,\quad R_x(\beta)=\left(\begin{array}{ccc}
\cos\beta  & 0 & -\sin\beta
\\
0 &  1&  0
\\
\sin\beta & 0  & \cos\beta
\end{array}
\right) .
\end{eqnarray*}

\begin{figure}
	\centering
    \includegraphics[width=\columnwidth]{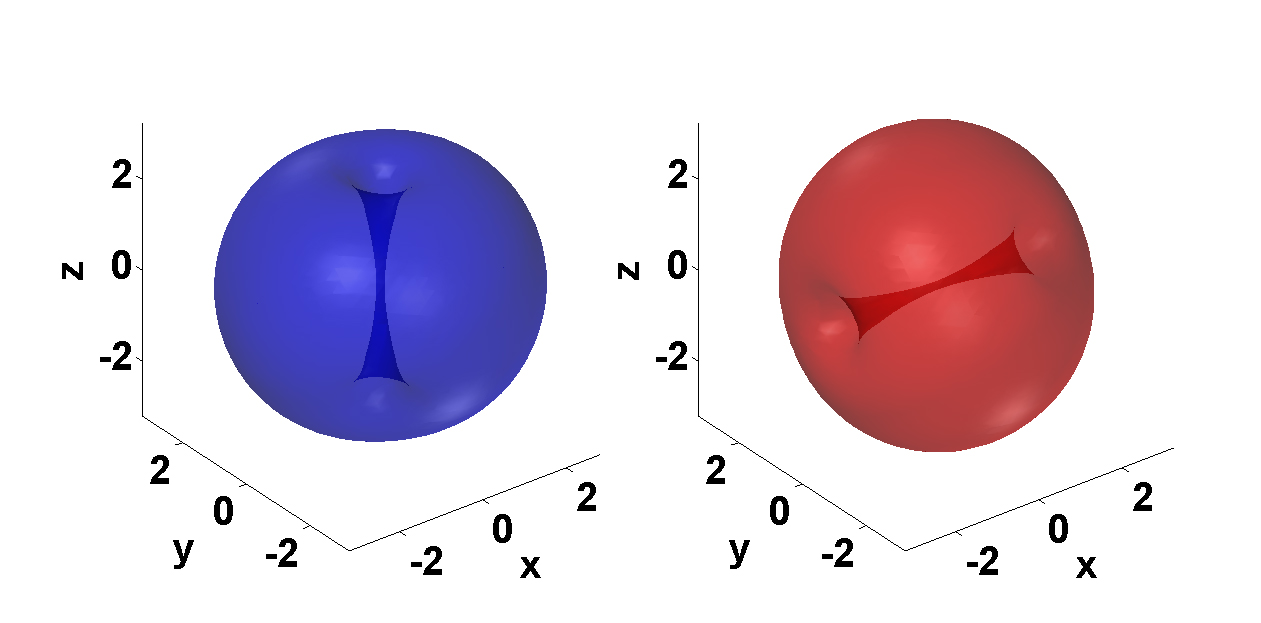}
	\includegraphics[width=\columnwidth]{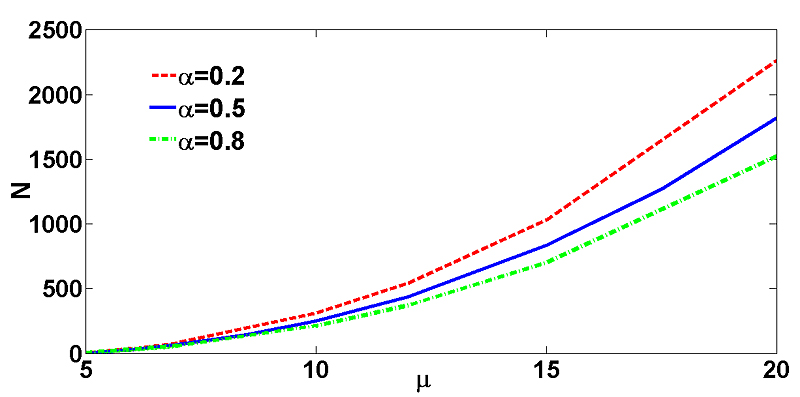}

	\caption{ (Color online) Isosurfaces of the components of the cross vortex corresponding to $\left\vert u_1 \right\vert ^{2}=1$ (left upper panel) and $\left\vert u_2 \right\vert ^{2}=1$  (right upper panel) for $\mu=10$  and dependence  $N$ {\it vs} $\mu$ for several values of the interspecies interactions.}
\label{fig1}
\end{figure}
%

Considering now the nonlinear family bifurcating from the spinor $(\tu_1,\tu_2)^T$ one can compute the interaction energy which is deposited into the system
when weakly nonlinearity is switched-on:
$
E_{int}(\beta)\approx \frac{\alpha}{2}\pi^{3/2}\left(\cos^2\beta+3 \right).
$
Thus $E_{int}(\beta)$ achieves its minimum at $\beta=\pi/2$, 
i.e. when the two vortex lines are mutually orthogonal, as illustrated in the upper panel of Fig.~\ref{fig1}.
A family of nonlinear cross vortices [characterized by the dependence $N(\mu)$] bifurcates from the state $(\tu_1,\tu_2)^T$ 
 corresponding to $\beta=\pi/2$.
The isotropy of the system with respect to the $y$ axis suggests that the entire family, i.e., the nonlinear cross vortices are also characterized by orthogonal vortex lines.
Such families are shown in the lower panel of Fig.~\ref{fig1} for several values of the interspecies interaction $\alpha$. It is evident from this figure that all these families grow from a single solution corresponding to $\mu=\tmu=5$.
Since there is no qualitative difference between the cases with different $\alpha$
we consider $\alpha=0.5$ hereafter.

\begin{figure}[t]
	\centering
	\includegraphics[width=\columnwidth]{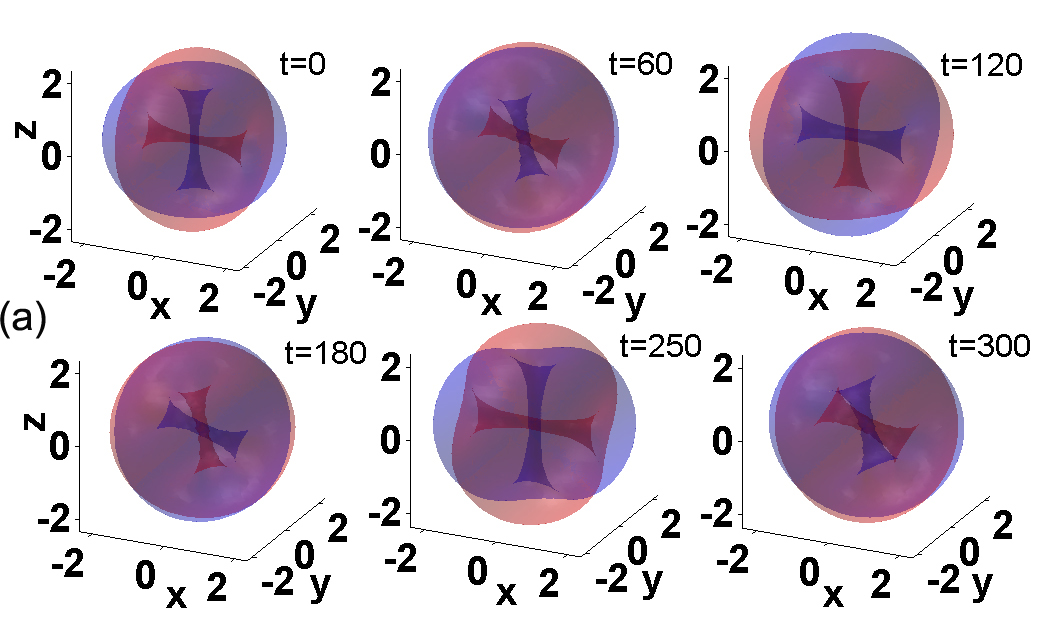}
	\includegraphics[width=\columnwidth]{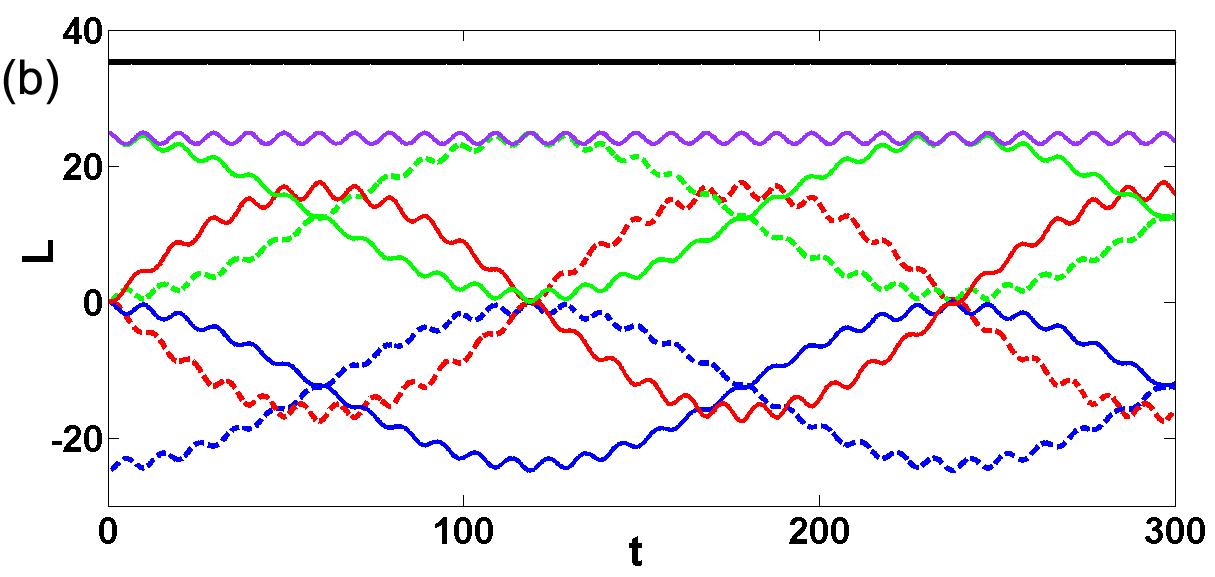}
	\includegraphics[width=0.9\columnwidth]{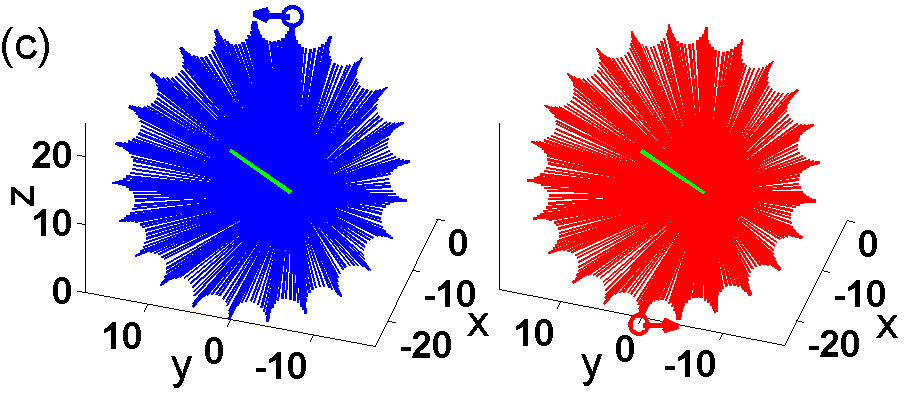}

	\caption{(Color online) (a) Snapshots of the cross vortex dynamics with N=25 at the instants of time indicated in the panels. The isosurfaces  correspond to $\left\vert \psi_{1,2}\right\vert ^{2}=0.1$. Blue and red colors represent the $\psi_1$ and $\psi_2$ components, respectively.
(b) Solid (dashed) lines show the dynamics of the $x$ (blue), $y$ (red), and $z$ (green) components
of the angular momentum $\bL_1$ ($\bL_2$).  The moduli of $\bL_{1}$ and $\bL_{2}$ are shown by
the overlapping purple lines
and the conserved modulus of the total angular momentum $\bL$ is shown by the solid black line.
(c) The evolution of $\bL_{1}$ (left panel) and $\bL_{2}$ (right panel).
The single green lines represent the invariant $\bL$. Circles with arrows indicate axes positions at $t=0$ as in upper left snapshot of panel (a).
}
\label{fig2}
\end{figure}

\begin{figure}
	\centering
	\includegraphics[width=8cm]{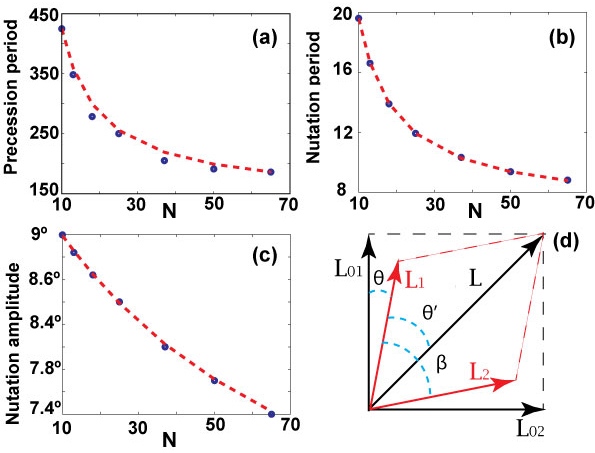}
	
	
	\caption{(Color online)  (a) Precession periods obtained numerically (circles) and interpolated by $T_{pr}^{(fit)}=142.6+2824.228/N-2.375/N^2$ (dashed line). (b) Nutation periods obtained numerically (circles) and interpolated by $T_{nut}^{(fit)}=3.5+160/N$ (dashed line). (c) Nutation amplitude obtained numerically (circles) and interpolated by $\theta^{(fit)}=9.5+0.056N-5.2\cdot 10^{-4}N^2-2.3\cdot 10^{-6}N^3$ (degrees, dashed line).  (b) Schematic diagram of the geometry of the angular momenta.}
	\label{fig3}
\end{figure}

Turning now to the dynamical problem we simulate numerically the evolution of a cross vortex
after an instantaneous switch-on of the nonlinearity.
We evolve the solution of (\ref{GPE}) starting from initial conditions
$\psi_1(t=0)=2.12(x+iy)e^{-r^2/2}$, $\psi_2(t=0)=2.12(z+iy)e^{-r^2/2}$ (the number of atoms in each component is $N_1=N_2\approx 25$).
Snapshots of the evolution of the binary condensate are shown in Fig.~\ref{fig2} (a).
We observe that both vortex lines experience persistent rotations returning to the initial state after the full period $T\approx 248$.
In Fig.~\ref{fig2}(b) we present the evolution of the
projections of $\bL_{1,2}$.
The dynamics of the angular momentum components shows two types of oscillations. The oscillations with a longer period and larger amplitude correspond to rotation of the individual momenta  $\bL_{1,2}$ with respect to the conserved total angular momentum $\bL$, i.e., the {\em precession} of the cross vortex. The precession is accompanied by oscillations with a smaller period and a smaller amplitude. The latter describe oscillations of $\bL_{1,2}$, whose amplitude is characterized by the angle $\theta$ measuring deviations of $\bL_{1,2}$ from their initial orientations $\bL_{j}(t=0)=\bL_{0j}$ [see Fig.~\ref{fig3}(d) below]. The dynamics of this second type corresponds to {\em nutation}. Fig.~\ref{fig2}(c) shows a vectorial representation of the evolution of the angular momenta
of each component $\bL_{1,2}$.

Our simulations demonstrate the robustness of the nonlinear cross vortices,
even though they are excited by initial conditions significantly deviating from stationary nonlinear
vortices. The dynamics is accompanied by  modulations of the vortex lines (and consequently of vortex shapes).
Numerical studies performed for different $N$ result
in qualitatively similar regimes, however,
the dynamical characteristics are significantly affected by the two-body  interactions.
This is illustrated in Fig.~\ref{fig3} where we show
the precession [Fig.~\ref{fig3}(a)] and nutation  [Fig.~\ref{fig3}(b)] periods
and the nutation amplitude [Fig.~\ref{fig3}(c)] {\em vs} the number of particles.
For small $N$ the precession and nutation are slow and they accelerate as $N$ increases,
while the nutation amplitude decreases.


\begin{figure}
  	\centering
    \includegraphics[width=7.5cm]{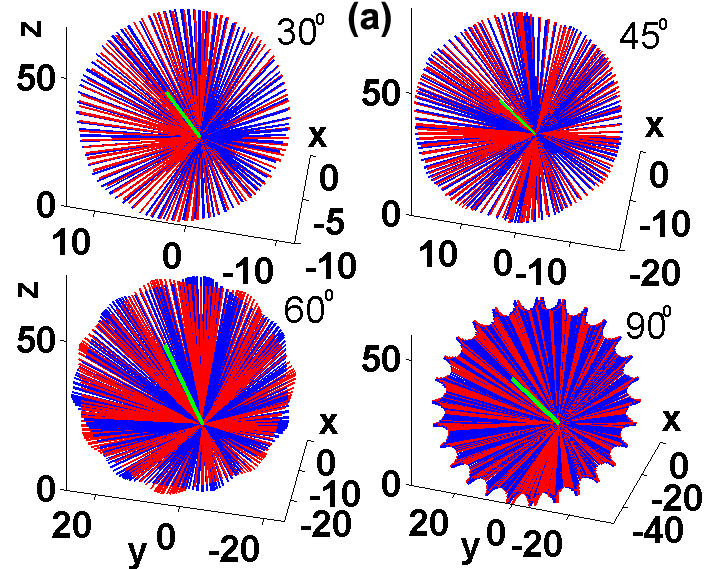}
    \includegraphics[width=8cm]{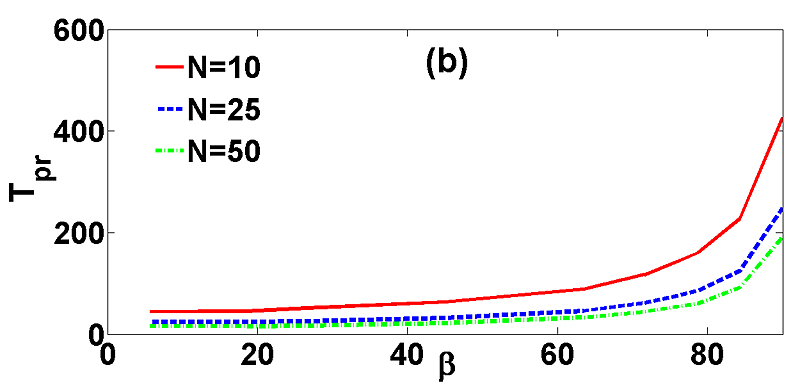}
  	\caption{ (Color online) (a) Precession of nonlinear vortices grown from the linear harmonics with $N=50.11$ and various values of $\beta$. (b) Dependence of the precession period on the initial angle $\beta$.}
  	\label{fig4}
  \end{figure}

The obtained dynamics is not obvious for at least two reasons.
First, earlier observed precession of vortices other types was explained by instabilities~\cite{SvFetter,VPG}.
In our simulations, however, we did not observe a transient period describing development of instabilities, but the rotational dynamics started immediately after switching-on the nonlinearity. On the other hand,  Eq.~(\ref{M12}) indicates that a "force" necessary to induce precession and nutation should be related to anisotropy of the atomic distributions in each of the vortices and cannot be induced at $t=0$ by the imposed initial conditions
[the right hand side of (\ref{M12}) is zero for any vortex of the form $\psi_j=(x+iy)\phi_j(r_\bot^2 ,z^2)e^{-i\mu t}$, where $\br_\bot=(x,y)$]. To understand the observed evolution we use a simple mechanical analogy shown in Fig.~\ref{fig3} (d). The angular momenta $\bL_{01}$ and $\bL_{02}$ of the linear modes used for the initial conditions, can be viewed as an initial perturbation of the angular momenta of the nonlinear cross-vortex with  $\bL_{1}$ and $\bL_{2}$. Since now $\bL_{1}\cdot\bL_{2}\neq 0$, i.e., $\beta(t=0)<\pi/2$, the dynamics shows a precession whose period is faster for a smaller $\beta(t=0)$ [see Fig.~\ref{fig4} (b)].
Concomitantly, due to the inter-atomic interactions there appears a "force" tempting to restore $\beta=\pi/2$ [i.e. to minimize $E_{int}(\beta)$]. This force is responsible for the nutation. We also verify that $\frac{d}{dt}(\bL_{1}\cdot\bL_{2})\neq 0$ (notice that the components of $\bL_1$ and $\bL_2$ oscillate out-of-phase) which means that the modula $L_{1,2}$ are not constant. 
We did not found stationary vortices with $\beta<\pi/2$, but only precessing ones [Fig.~\ref{fig4} (b)]. This means that $\beta<\pi/2$ defines "initial" angular velocity of the precession ($\beta=\pi/2$ corresponds to zero velocity). This difference in the initial angular velocities is clearly visible when comparing panels showing the dynamics in Fig.~\ref{fig4} for different initial angles.

Both precession and nutation disappear in the linear limit ($N\to 0$): the periods become infinite.
Thus $1/N$ can be viewed as a small parameter for the estimate of the periods.
Figs.~\ref{fig3}(a) and (b) show the respective fitting curves.
The dependence of the angular nutation amplitude is approximated by an
almost linear dependence on $N$, see Fig.~\ref{fig3}(c).

Although, the presented "mechanical" picture is over-simplified, it results in reasonable estimates for the dynamical parameters. If $\langle L_j\rangle$ is the average value of $|\bL_j|$, the average angle of precession can be estimated as $\theta^\prime=\arccos(2\langle L_1\rangle /L)$ and the average angle of nutation as $\theta=\pi/4-\theta^\prime$, respectively.
Thus the ratio between the nutation and precession periods can be estimated as $T_{nut}/T_{pr}=\sin(\theta^\prime)/\sin(\theta)$.
Using these simple estimates for the case of $N=25$ shown in Fig.~\ref{fig2}, we obtain  $\theta^\prime\approx 0.75196$, $\theta\approx 0.03344$ and hence $T_{nut}/T_{pr}\approx 20.42$.
The result of our simple considerations is not very far from the numerically observed value of $\left(T_{nut}/T_{pr}\right)_{num}\approx 25.05$ but clearly deviates from it.
This difference can be explained by the strength of the inter-atomic interaction which results in a significantly perturbed nonlinear dynamics.
Reducing the number of particles to $N=10$ (not shown here), i.e., reducing the influence of the nonlinearity,
one obtains much "cleaner" precession and nutation.
For this case, our estimate yields  $\theta^\prime\approx 0.7537$, $\theta\approx 0.0317$ and hence $T_{nut}/T_{pr}\approx 21.6$
while numerically we find $\left(T_{nut}/T_{pr}\right)_{num}\approx 21.9$
which is a remarkable accuracy.

  %
Remarkably, a similar dynamics is also observed for initial conditions
with arbitrary initial angles between the angular momenta $\bL_1$ and $\bL_2$.
We start again from the linear 3D harmonics oriented this time arbitrary with respect to each other and instantaneously switch-on the nonlinear interaction.
Examples of the vortex precession and nutation for different initial $\beta$ in (\ref{R}) are shown in Fig.~4(a).  The dynamics obtained for $\beta< \pi/2$ degrees
are clearly different from those for $\beta = \pi/2$ degrees since for $\beta\neq \pi/2$ instead of spikes
smooth oscillations appear which correspond to a precession with a forward release speed ~\cite{precessions}.
For small $\beta$ the amplitude of the nutations decreases significantly and is
hardly observable. Also for smaller values of $\beta$ precession is much faster, therefore the opening angles of axes cones decrease with $\beta$ as we can see comparing the scales of panels of Fig. 4(a) .
Fig.~\ref{fig4}(b) shows the dependence of the full precession period on the initial angle $\beta$ (for the sake of compactness we merged motion of two axes into single panels showing evolution of axes of $\psi_1$ and $\psi_2$ with blue and red colors respectively). For all considered $N$ the period decreases strongly when $\beta$ deviates from $\pi/2$.

To conclude, we have reported that a two-component
BEC with repulsive inter- and intra-species interactions
can support very robust  non-coaxial vortices, with approximately orthogonal vortex lines
in each of the components. These cross vortices can be excited from noninteracting eigenstates of the 3D linear harmonic oscillator by sudden switch-on the nonlinearity and they are characterized by remarkably persistent dynamical regimes of precession with nutation resembling the motion of a rigid body.
The precession is generated by the initial conditions, while the nutation occurs due to the force caused by the inter-species interactions and is associated with deformations of vortex shapes.



\begin{acknowledgements}

RD and TM acknowledge support
by the Deutsche Forschungsgemeinschaft (DFG) via the
Research Training Group (GRK) 1464 and computing time provided by PC$^2$
(Paderborn Center for Parallel Computing).
RD
acknowledges support by the Russian
Federation Grant 074-U01 through ITMO Early Career Fellowship scheme.
VVK acknowledges support of the FCT (Portugal) under the grant UID/FIS/00618/2013.

\end{acknowledgements}

\end{document}